\documentstyle[prl,aps,multicol,psfig]{revtex}

\begin{document}
\draft
\preprint{}
\title{Finite size scaling in neural networks}

\author{Walter Nadler}
\address{Institut f\"ur Theoretische Chemie, Universit\"at T\"ubingen,
Auf der Morgenstelle 8, D-72076 T\"ubingen, Germany
}
\author{Wolfgang Fink}
\address{Institut f\"ur Theoretische Physik, Universit\"at T\"ubingen, 
Auf der Morgenstelle 14, D-72076 T\"ubingen, Germany
}

\date{\today}

\maketitle

\begin{abstract}

We demonstrate that the fraction of pattern sets that can be stored
in single- and hidden-layer perceptrons exhibits finite size scaling.
This feature allows to estimate the critical storage capacity $\alpha_c$
from simulations of relatively small systems. We illustrate this approach
by determining $\alpha_c$, together with the finite size scaling exponent $\nu$,
for storing Gaussian patterns in committee and parity
machines with binary couplings and up to $K=5$ hidden units.

\end{abstract}


\begin{multicols}{2}

\narrowtext

Finite size scaling (FSS) has proven to be a powerful method for analyzing 
phase transitions, which occur rigorously only in the thermodynamic limit, 
using simulations of systems of finite size \cite{Privman90}. In particular,
it has become the prime method for determining numerical
values of critical coupling parameters and exponents \cite{SA92}.

Phase transitions are known to occur not only in condensed matter \cite{DG}
and percolation systems \cite{SA92},
but also in random graphs \cite{Palmer85}, neural networks \cite{HKP91},
and in algorithmic problems like search \cite{WH93} and 
the satisfiability of random boolean expressions \cite{MSL92}.
Heuristic derivations of FSS rely on the divergence
of a correlation length at a critical point in the infinite system \cite{SA92,KS85}.
However, Kirkpatrick and Selman \cite{KS94} have demonstrated recently that 
FSS can be used efficiently also in problems 
without any intrinsic length scales, like the
connectivity of random graphs and the satisfiability of random boolean expressions.
Abstract neural networks \cite{HKP91} are another class of systems without
intrinsic length scale, and we will show in this contribution
that FSS occurs at the transition from storable to unstorable pattern set sizes,
and that it provides a powerful computational method for determining critical storage 
capacities.

\begin{figure}

\psfig{figure=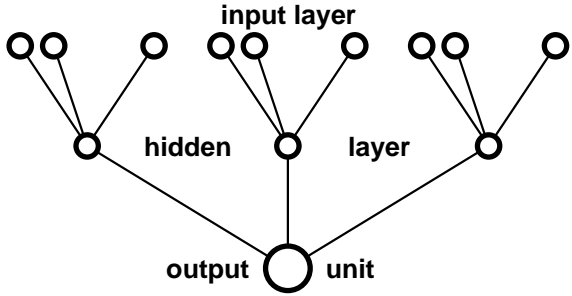,width=3truein}

\caption{Tree-like multi-layer perceptron with $K=3$ hidden units.}
\label{figure1} 
\end{figure}

We will concentrate on particular feed-forward networks of the perceptron class, 
namely multi-layer perceptrons with $N$ input neurons,
$K$ hidden units, and a regular tree-like connectivity
($N\text{ mod }K=0$), see Fig.~\ref{figure1}, 
which are also known as $committee$ and $parity$ machines
(CM, PM) with non-overlapping receptive fields \cite{MD89,BHK90,BHS92}.
Input patterns $\xi_{ik}$, $k=1,\ldots,K$, $i=1,\ldots,N/K$,
are processed by the following rules:
The output of hidden layer cell $k$ is given by
\begin{equation}
\label{ioeq1}
O_k = \text{sgn} \left(\sum_{i=1}^{N/K} J_{ik}\xi_{ik} \right) \quad ,
\end{equation}
$J_{ik}$ being the coupling between input cell $ik$ and
hidden unit $k$,  while the final output is determined by
\begin{equation}
\label{ioeq2}
O = \text{sgn} \left(\bigodot_{k=1}^K O_k \right) \quad ,
\end{equation}
where in the case of a CM the majority rule is implemented
by $\bigodot\equiv\sum$,
while in the case of a PM $\bigodot\equiv\prod$.
A standard single-layer perceptron corresponds to $K=1$. 
Since the majority rule is somewhat problematic in case of even $K$, 
we will restrict ourselves here to CM with $K$ odd.

A perceptron is able to store a particular
set of input patterns $\{\xi_{ik}^\mu\}$, $\mu=1,\ldots,p$,
if there exists a coupling set $\{J_{ik}\}$ such that -
under the action of Eqs.~(\ref{ioeq1},\ref{ioeq2}) -
a prescribed set of outputs $\{O^\mu\}$ is generated.
It is well known that for small values of $\alpha = p/N$
such a set of couplings can always be found, while for large
enough $\alpha$ the probability for its existence vanishes.
For finite systems the fraction of all possible 
input-output relations $\{(\xi_{ik}^\mu,O^\mu)\}$ of relative size
$\alpha$ that can be stored, which we will call $P(\alpha,N)$ \cite{note1}, 
undergoes a smooth transition from one to zero. However,
in the infinite system it switches from one to zero at the 
critical storage capacity $\alpha_c$.

\begin{figure}

\psfig{figure=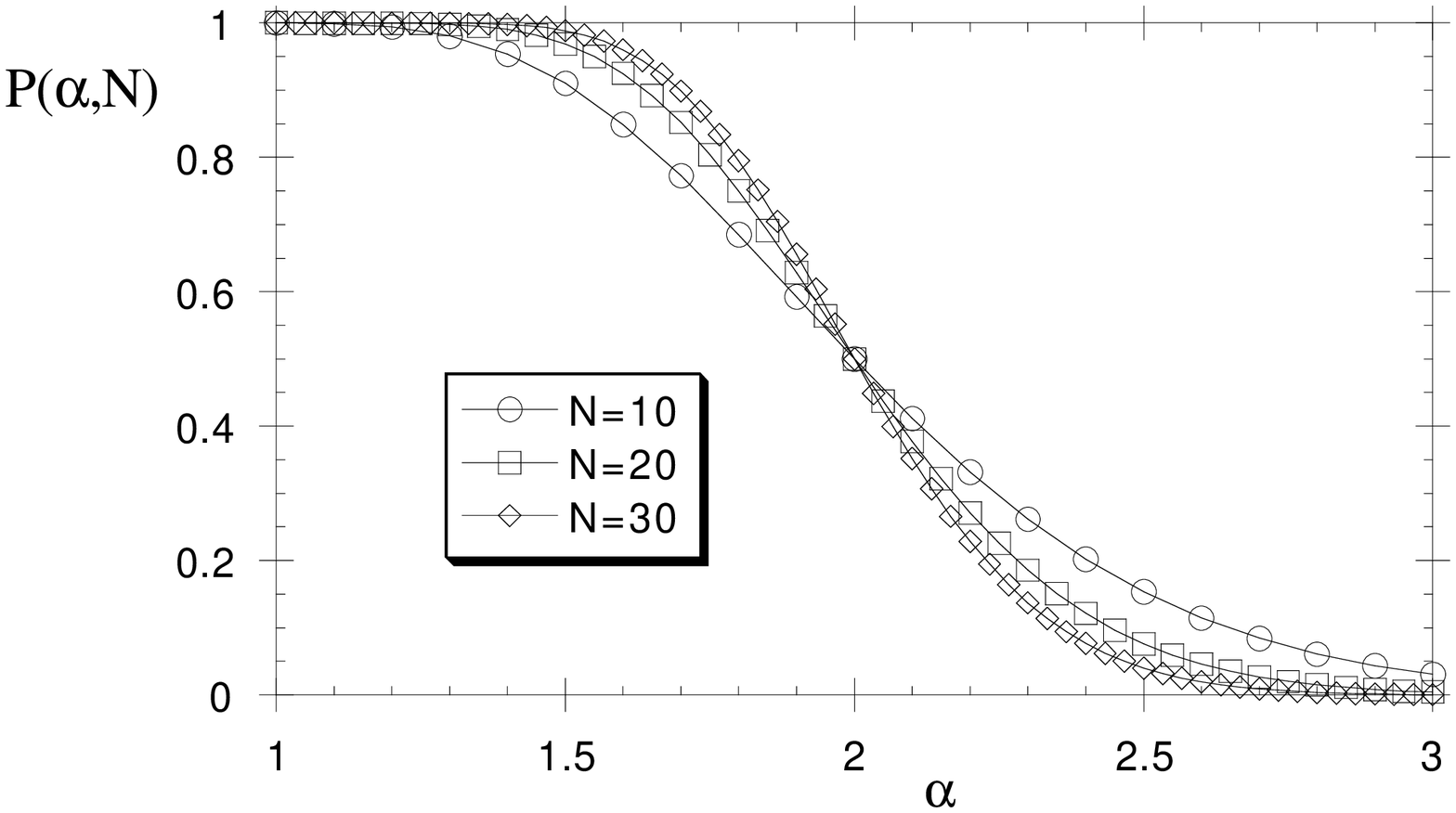,width=4truein}

\psfig{figure=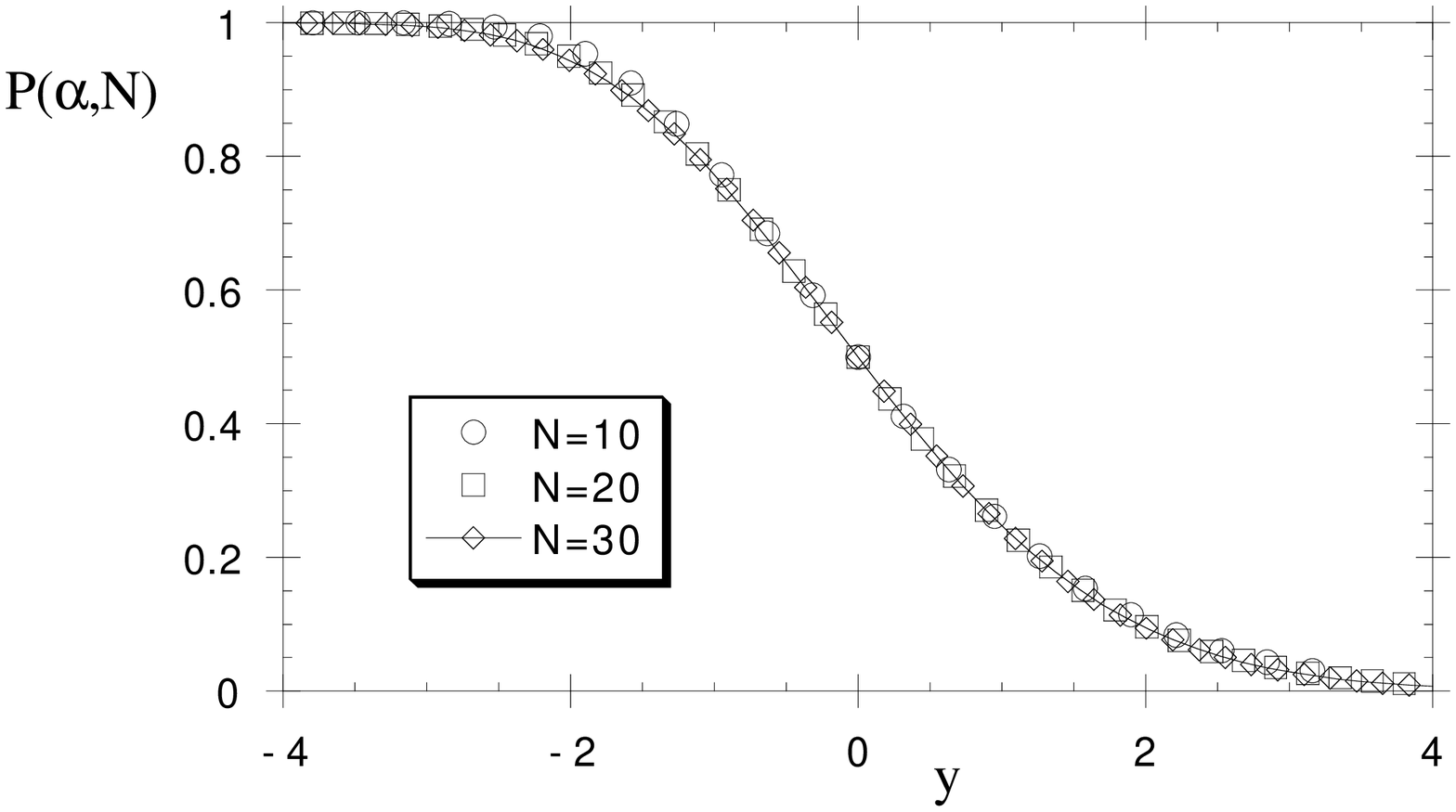,width=4truein}

\caption{Finite size scaling in the 
single-layer perceptron with continuous couplings:
(top) Eq.~(\protect\ref{CoverSolutionEquation}),
(bottom) finite size scaling as indicated in the text.
}
\label{CoverSolutionPlot} 
\end{figure}

This behavior, together with FSS, is nicely illustrated for the single-layer perceptron
with continuous couplings and
the $\xi_{ik}$ drawn from a Gaussian distribution,
where the exact solution for 
$P(\alpha,N)$ is known analytically\cite{HKP91,Cover65},
\begin{equation}
\label{CoverSolutionEquation}
P(p/N,N) = 2^{1-p}\sum_{i=0}^{N-1} { p -1 \choose i} \quad .
\end{equation}
Figure~\ref{CoverSolutionPlot} (top) shows $P(\alpha,N)$
for various values of $N$. The common intersection of these 
curves at $\alpha=2$ is noticed immediately.
Also, the steepness of the transition increases with system size $N$.

Under FSS, systems of different size behave
in an identical way near the transition
under a size-dependent rescaling of the control parameter \cite{KS94},
\begin{equation}
\label{scalingEquation}
y = (\alpha-\alpha_c)N^{1/\nu} \quad .
\end{equation}
Necessarily, the common intersection of the transition curves 
observed above corresponds to the critical storage capacity $\alpha_c$.
Figure \ref{CoverSolutionPlot} (bottom) shows that a rescaling with $\nu=2$
and $\alpha_c=2$
indeed lets all transition curves fall onto a single scaling curve.
In this particular case, 
the numerical value of the FSS exponent $\nu$, together with 
the analytic form of the scaling function, 
\begin{equation}
\label{scalingFunctionK1}
f(y) = {1\over2} + {1\over2} \text{erf}\left(-y/2\right)
\end{equation}
can be derived from the asymptotic behavior of 
Eq.~(\ref{CoverSolutionEquation}),
\begin{equation}
\label{asymptoticsK1}
P(\alpha,N) \to {1\over2} + {1\over2} 
\text{erf}\left(\sqrt{N\over2\alpha}\left(2-\alpha\right)\right) \quad .
\end{equation}
Figure~\ref{CoverSolutionPlot} demonstrates moreover that critical storage capacity
$\alpha_c$ and FSS exponent $\nu$ can already be estimated from systems
of relatively small size.

Simulations of neural networks are plagued by the problem
that learning algorithms \cite{HKP91}, necessary to determine coupling sets that
solve the storage problem, are not guaranteed to reach a solution practically,
i.e. under realistic time constraints, even if it exists. Close to $\alpha_c$ the 
average learning time
diverges \cite{PrielEtAl94}, a behavior reminding of critical slowing down \cite{DG}.
The situation is worse for systems with binary couplings,
since there the usual learning algorithms are not applicable 
\cite{Koehler90,Horner92,Horner93,Patel93}.

We will concentrate in the following on perceptrons with binary couplings
$J_{ik}=\pm1$, also known as Ising perceptrons.
Employing complete enumeration of the couplings for systems up to size $N=30$,
simulation results $independent$ of any learning algorithm are obtained.
We used Gaussian patterns for the results presented in this contribution.
Note that for binary coupling perceptrons with a finite number of hidden units
information theory gives an upper limit for the critical storage capacity of one,
i.e. $\alpha_c\le1$ \cite{BHS92,GD88}.

\begin{figure}

\psfig{figure=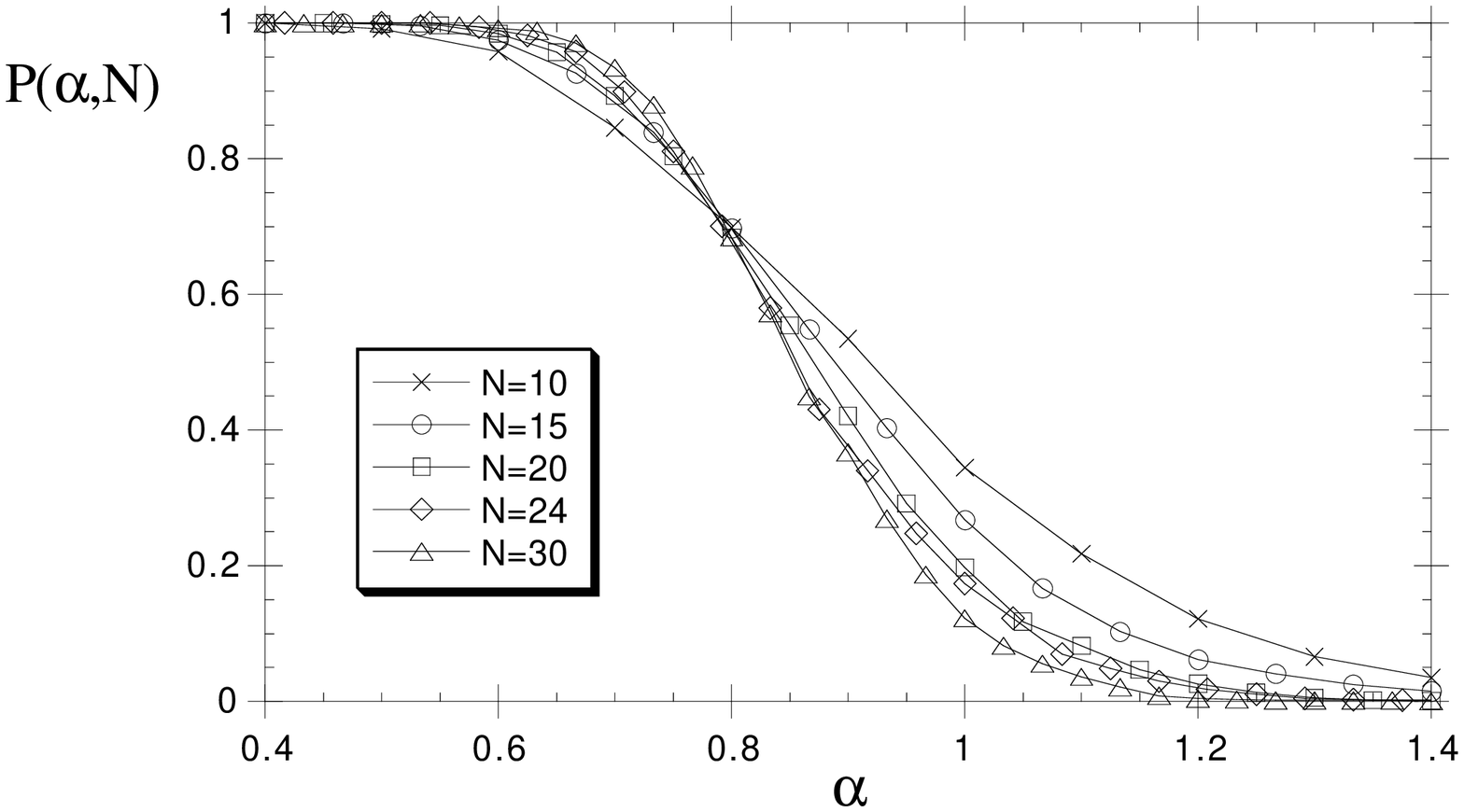,width=4truein}

\psfig{figure=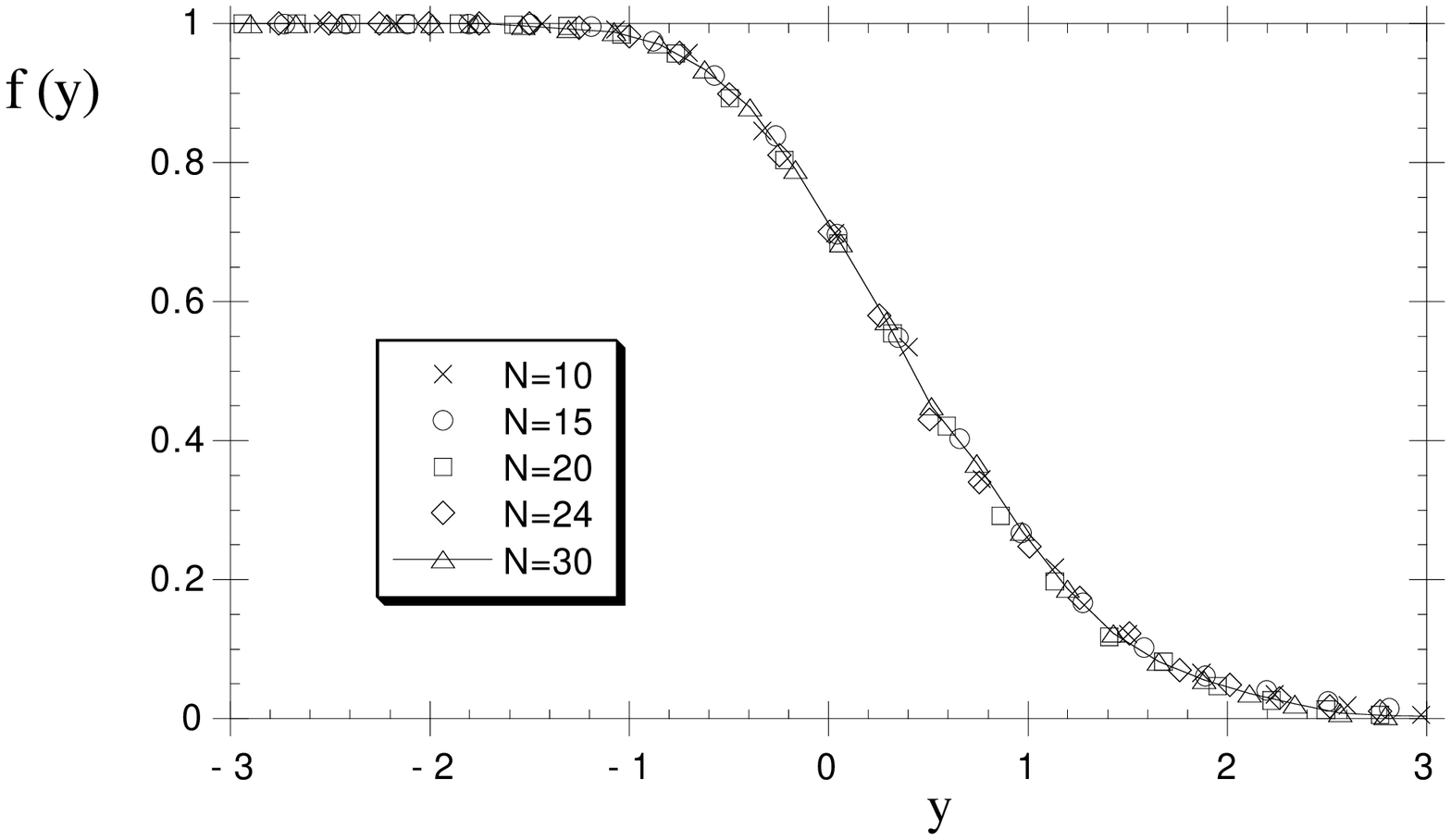,width=4truein}

\caption{Finite size scaling in the single-layer perceptron
with binary couplings: (top) before,
(bottom) after finite size scaling as indicated in the text.
}
\label{K1Plot}
\end{figure}

Figure \ref{K1Plot} (top) shows simulation results for $P(\alpha,N)$ for the 
case of a single-layer binary coupling perceptron. 
Sets of input-output relations
were classified as storable or unstorable by complete enumeration
of the coupling space \cite{note2}.  
Each data point was sampled with about $10^3$ randomly chosen 
sets of input-output relations, giving a 
relative error of about $3\%$. As in the case of continuous couplings, 
Fig.~\ref{CoverSolutionPlot}, the curves for various system sizes intersect 
at the critical storage capacity, here with the numerical value
$\alpha_c\approx0.8$.
Figure \ref{K1Plot} (bottom) shows the same data under rescaling with
Eq.~(\ref{scalingEquation}) and $\nu\approx1.7$. Again, all data points fall onto
one scaling curve. Note that the value of the scaling function at the transition,
 $f(0)\approx0.7$, is different from the continuous case ($f(0)=0.5$).

Results for the hidden-layer systems of parity and committee type show 
a behavior qualitatively similar to the one presented in  
Fig.~\ref{K1Plot} for the single-layer perceptron. 
We have collected our results for various values of $K$
in Table \ref{table}. As it is to be expected,
$\alpha_c$ increases with the introduction of a hidden layer of neurons.
The {FSS} exponent $\nu$ decreases with increasing $K$, to about 1.3 and 1.2 for CM,
and to values around one for PM.

The most surprising results are those for PM.
Already a system with $K=2$ hidden units exhibits a
storage capacity extremely close to the theoretical limit,
and Table \ref{table} shows that there is practically no improvement 
in increasing $K$. 
In application situations, storing patterns
has to be done using finite size perceptrons.
Since the \hyphenation{FSS} FSS scaling function $f(y)$ 
describes the asymptotic behavior of the
fraction of storable patterns, $P(\alpha,N)$,
around $\alpha_c$,
the critical capacity has to be considered
together with $f(y)$  
when assessing the quality of a particular system.
Note that $f(y)$ decreases considerably with $K$ 
in the critical region for PM as well as CM, 
see $f(0)$ in Table~\ref{table}.
These features
suggest that a PM with $K=2$ is already the best $practical$ binary perceptron
for storing continuous patterns.

Simulation studies of the single-layer binary perceptron
have been performed before for the problems of storing binary 
\cite{Koehler90,Horner92,Horner93,Patel93,DGP91,KS93}, 
and Gaussian patterns \cite{DGP91,KO89}, 
using various approaches and not always leading to conclusive results. 
Our result for $\alpha_c$ differs significantly from the analytical result of Ref.~\cite{KM89} 
($\alpha_c=0.833$) obtained using a first order replica symmetry breaking ansatz (RSB),
but could be considered compatible - within error bars - 
with the simulation result of Ref.~\cite{KO89} ("$\alpha_c\approx0.82$" \cite{ErrorBars}).
This discrepancy between the analytical approximation
and our simulation result suggests - 
provided finite size scaling holds -
that the first order RSB is still insufficient
for a correct analytical treatment of the $K=1$ case,
despite the claims in \cite{KM89}.
For binary CM and PM storing Gaussian patterns
no analytical or simulation results are available
at present, to the best of our knowledge. 

It has been hypothesized on the basis of replica studies \cite{KO89} 
that the storage capacity for binary and Gaussian patterns is identical. 
Previous simulation results for $K=1$ seemed to be compatible with this hypothesis and 
with the RSB result reported above
($\alpha_c=0.83$ \cite{Koehler90}, $\alpha_c=0.833$ \cite{Horner92,Horner93,Patel93},
however \cite{ErrorBars}).
Since our results differ significantly from the RSB result,
this casts some doubt on either this hypothesis, the RSB result,
or on the interpretation of the simulation results \cite{ErrorBars}.
For the case of storing binary patterns in CM,
simulation results using complete enumeration
have been obtained for $K=3$ in \cite{BHS92},
together with analytical results 
for $K=3$ ("$\alpha_c\approx0.92$"), and for $K\to\infty$ ("$\alpha_c\approx0.95$"), 
using a replica symmetric (RS) ansatz.
Although our simulation results for CM differ somewhat, they can still
be considered statistically compatible with those values, 
in contrast to the $K=1$ case discussed above. 
This result supports the hypothesis of \cite{BHS92} that a RS ansatz might 
be sufficient for CM, and suggests that the hypothesis of an identical 
$\alpha_c$ for storing binary and Gaussian patterns might hold at least for CM.

In closing, we like to draw again attention to the fact that the values for $f(0)$
differ strongly between various perceptrons.
In particular, with the single exception of the  CM with $K=3$, they differ 
considerably from $1/2$.
On the other hand, the relation $P(\alpha_{0.5}(N),N)=0.5$ has 
often been the basis of an extrapolation to the 
infinite system critical parameter from simulations of finite systems
\cite{EK93,KS93}.
If we define $y_{0.5}$ by $f(y_{0.5})=0.5$, then 
\begin{equation}
\alpha_{0.5}(N)=\alpha_c+y_{0.5}N^{-1/\nu} \quad .
\end{equation}
Together with the fact that the FSS exponent $\nu$ deviates from one 
particularly for $K=1$ and for CM, this feature emphasises the need for an 
extrapolation nonlinear, instead of linear, in $1/N$ to correctly
obtain the thermodynamic limit value of $\alpha_{0.5}(N)$ \cite{KS94Corr},
and it may be the source of some problems encountered
in earlier simulation studies \cite{DGP91,KS93}.

The above results demonstrate that the FSS ansatz not only offers a new 
and powerful computational approach for evaluating the critical storage capacities
of binary perceptrons, but also allows a detailed view on the storage properties
in the critical region. We believe that it will prove valuable in analyzing the 
properties of a wide variety of binary perceptron topologies.

\begin{table}
\caption{critical storage capacity $\alpha_c$, 
finite size scaling exponent $\nu$, and 
transition value $f(0)$ of scaling function $f$,
for various binary perceptrons\protect\tablenotemark[1].}
\label{table}
\begin{tabular}{lddd}
K & $\alpha_c$ (SD) & $\nu$ (SD) & $f(0)$ (SD)\\
\tableline
\multicolumn{4}{c}{\it single-layer}\\
1   & 0.796 (0.010) &  1.68 (0.09)  &   0.70 (0.03)\\
\multicolumn{4}{c}{\it committee machine}\\
3   &    0.899 (0.008)  & 1.28 (0.06)  &   0.49 (0.03) \\
5   &    0.932 (0.012)  & 1.15 (0.08)  &   0.36 (0.04) \\
\multicolumn{4}{c}{\it parity machine}\\
2   &    0.992 (0.005)  & 1.02 (0.04)  &   0.37 (0.02) \\
3   &    0.998 (0.005)  & 0.93 (0.03)  &   0.22 (0.02) \\
4   &    0.999 (0.008)  & 0.97 (0.04)  &   0.12 (0.02) \\
5   &    0.983 (0.009)  & 0.91 (0.04)  &   0.07 (0.01) \\
\end{tabular}
\end{table}
\tablenotemark[1]{In order to perform a reproducible
and unambigious error analysis of the data we used the {\it bootstrap} method 
\protect\cite{ET93}: 
About $10^3$ bootstrap samples were drawn from the original data for all system sizes,
and for each such sample $\alpha_c$ and $\nu$
were determined together by minimizing
the mutual mean squared deviation of the interpolating scaling curves; 
the presented values and the estimated errors 
are the means and standard deviations, respectively,
of $\alpha_c$, $\nu$, and $f(0)$ in the set of bootstrap samples.
}

\end{multicols}


\begin{references}
\bibitem{note1}
We note that $P(\alpha,N)$ --- being the probability to find, in the ensemble of 
all sets of input-output relations of relative size $\alpha$, 
a set of relations that can be stored in a perceptron of size $N$ --- 
differs considerably from the coupling space volume 
$V(\alpha,N)$ that is usually determined in analytical calculations, 
e. g. based on the replica approach ( see for example Ref. \protect\cite{HKP91}). 
The latter is a fluctuating quantity, and care has to be taken
to its ensemble averaging ( $\overline{\ln(V)}$ vs $\ln(\overline{V})$ ).
This does not apply to $P(\alpha,N)$.
Moreover, $P(\alpha,N)$ is actually the more fundamental 
observable for the storage problem \protect\cite{HKP91}. 
However, lack of the ability to calculate
it analytically for many interesting systems
has lead to some neglect of this important observable
in the current literature.
\bibitem{ErrorBars}
Note that most previous simulation results
are distinguished by a complete lack of an estimate
for the error due to Monte Carlo sampling
of the extrapolated value for $\alpha_c$. On the other hand, the minuscule
error reported in \protect\cite{Patel93} (even smaller than that for several
individual data points used in the extrapolation), is hard to believe. 
Note also that, when learning algorithms are employed in simulations,
systematic errors may arise,
and further analysis has to be supplied with some theory on its
performance for large system sizes \protect\cite{Horner92,Horner93}.
Otherwise the choice of what system sizes to include in extrapolations
becomes somewhat arbitrary (as in \protect\cite{Koehler90,Patel93}).
\bibitem{KS94Corr}
This problem was noted already in Ref.~\protect\cite{KS94}.
\end{references}
\end{document}